\documentclass[preprint,showpacs,preprintnumbers,superscriptaddress,amsmath,amssymb]{revtex4}
\usepackage{graphicx}
\usepackage{dcolumn}
\usepackage{bm}
\usepackage[american]{babel}
\begin{document}

\title{New PVLAS results and limits on magnetically induced optical rotation and ellipticity in vacuum}

\author{E.~Zavattini\footnote{Deceased January 9, 2007}}
\affiliation{INFN, sezione di Trieste and Dipartimento di Fisica, Universit\`a di Trieste, Trieste, Italy}
\author{G.~Zavattini}
\affiliation{INFN, Sezione di Ferrara and Dipartimento di Fisica, Universit\`a di Ferrara, Ferrara, Italy}
\author{G.~Ruoso}
\affiliation{INFN, Laboratori Nazionali di Legnaro, Legnaro, Italy}
\author{G.~Raiteri}
\affiliation{INFN, sezione di Trieste and Dipartimento di Fisica, Universit\`a di Trieste, Trieste, Italy}
\author{E.~Polacco}
\affiliation{INFN, Sezione di Pisa and Dipartimento di Fisica, Universit\`a di Pisa, Pisa, Italy}
\author{E.~Milotti}
\affiliation{INFN, sezione di Trieste and Dipartimento di Fisica, Universit\`a di Trieste, Trieste, Italy}
\author{V.~Lozza}
\affiliation{INFN, sezione di Trieste and Dipartimento di Fisica, Universit\`a di Trieste, Trieste, Italy}
\author{M.~Karuza}
\affiliation{INFN, sezione di Trieste and Dipartimento di Fisica, Universit\`a di Trieste, Trieste, Italy}
\author{U.~Gastaldi}
\affiliation{INFN, Laboratori Nazionali di Legnaro, Legnaro, Italy}
\author{G.~Di Domenico}
\affiliation{INFN, Sezione di Ferrara and Dipartimento di Fisica, Universit\`a di Ferrara, Ferrara, Italy}
\author{F.~Della Valle}
\affiliation{INFN, sezione di Trieste and Dipartimento di Fisica, Universit\`a di Trieste, Trieste, Italy}
\author{R.~Cimino}
\affiliation{INFN, Laboratori Nazionali di Frascati, Frascati, Italy}
\author{S.~Carusotto}
\affiliation{INFN, Sezione di Pisa and Dipartimento di Fisica, Universit\`a di Pisa, Pisa, Italy}
\author{G.~Cantatore}
\affiliation{INFN, sezione di Trieste and Dipartimento di Fisica, Universit\`a di Trieste, Trieste, Italy}
\author{M.~Bregant}
\affiliation{INFN, sezione di Trieste and Dipartimento di Fisica, Universit\`a di Trieste, Trieste, Italy}
\collaboration{PVLAS Collaboration}
\noaffiliation

\date{\today}

\begin{abstract}
In 2006 the PVLAS collaboration reported the observation of an optical rotation generated in vacuum by a magnetic field. To further check against possible instrumental artifacts several upgrades to the PVLAS apparatus have been made during the last year. Two data taking runs, at the wavelength of 1064 nm, have been performed in the new configuration with magnetic field strengths of 2.3 T and 5~T. The 2.3~T field value was chosen in order to avoid stray fields. The new observations do not show the presence of a rotation signal down to the levels of  $1.2\cdot 10^{-8}$ rad at 5~T and $1.0\cdot 10^{-8}$ rad at 2.3 T (at 95\% c.l.) with 45000 passes in the magnetic field zone. In the same conditions no ellipticity signal was detected down to $1.4\cdot 10^{-8}$ at 2.3 T (at 95\% c.l.), whereas at 5~T a signal is still present. The physical nature of this ellipticity as due to an effect depending on $B^2$ can be excluded by the measurement at 2.3~T. These new results completely exclude the previously published magnetically induced vacuum dichroism results, indicating that they were instrumental artifacts. These new results therefore also exclude the particle interpretation of the previous PVLAS results as due to a spin zero boson. The background ellipticity at 2.3~T can be used to determine a new limit on the total photon-photon scattering cross section of $\sigma_{\gamma\gamma} < 4.5 \cdot10^{-34}$ barn at 95\% c.l..
\end{abstract}


\maketitle

\section{Introduction}

Non linear effects in electromagnetic processes in vacuum have been sought after for many years after having been predicted by Euler and Heisenberg in their effective Lagrangian published in 1936 \cite{QED}. The only input to their calculation was the Heisenberg uncertainty principle leading to virtual pair creation, which allowed photons to interact with each other. The direct measurement of this effect is yet to be seen and has been the aim of the PVLAS experiment since its beginnings. The PVLAS experiment \cite{Bakalov}, financed by the Italian Istituto Nazionale di Fisica Nucleare (INFN), is located at the Laboratori Nazionali di Legnaro of INFN, Padova, Italy. The setup consists of a  sensitive ellipsometer attempting to detect the small changes in the polarization state of light propagating through a 1 m long magnetic field region in vacuum. It is based on a high finesse Fabry-Perot cavity and a  superconducting 5~T rotating dipole magnet. Indeed, vacuum will become birefringent in the presence of a strong magnetic field \cite{Adler, Iacopini}. A possibile secondary effect, which could mask the vacuum magnetic birefringence, could be due to the existence of a light, neutral pseudoscalar/scalar particle coupling to two photons via the Primakoff effect \cite{Anselm, Maiani, Raffelt, Mass˜}. During a number of data taking campaigns from 2000 to 2005, the PVLAS collaboration systematically observed both an induced ellipticity and a rotation which were acquired by an initially linearly polarized laser beam after having traversed a 5 T magnetic field in vacuum \cite{PVLASrot, ell_talk}. These observations were at variance with the Euler-Heisenberg effective Lagrangian predictions in that the observed ellipticity was about $10^4$ times greater than expected. Furthermore, a rotation was observed which was not predicted. If one interpreted the observations as due to the existence of a light, neutral, spin-zero boson and used the results previously obtained by the BFRT experiment \cite{Cameron}, the values for mass and inverse coupling of $m \approx 1$~meV and $M \approx 4\cdot 10^5$ GeV, respectively, were found. These values, however, are in strong contradiction with the results from the CAST experiment \cite{Cast} and with other astrophysical bounds \cite{astro_bounds}. Many theoretical papers attempting to reconcile the CAST and PVLAS observations were published \cite{Masso:2005ym} and several ``photon-regeneration" experiments were started \cite{RegVanBibber, ph_reg_exp} to try to directly detect the particle candidate in an appearance experiment rather than in a disappearance one, as it is the case in the PVLAS experiment.
The published PVLAS rotation results regarded an empirical finding which was attributed to an effect originating in the Fabry-Perot cavity with the magnetic field energized. The origin of this signal, whether physical or instrumental, was unknown. However, the diagnostic tests originally performed allowed one to localize the effect in the cavity and to exclude several spurious signal sources, such as those due to electromagnetic pick-ups or to a direct action of the magnetic fringe fields on the optical components. In fact, given that it was not possible to completely eliminate them, fringe fields remained a plausible source of instrumental artifacts, albeit in conjunction with some yet to be found indirect effect. After a series of apparatus upgrades designed to minimize the effect of the fringe fields, which is discussed below, several measurement runs were carried out  both at the field strength of 5~T and at the reduced field intensity of 2.3 T, when the stray field intensity drops from 2--3 G (at a 5 T central field) down to 30--40 mG. The results from these measurements do not confirm the presence of a rotation signal at the expected frequency, also excluding the presence of an ellipticity signal at 2.3 T. The details of these measurements are discussed below. The background ellipticity and rotation values can be used to establish upper bounds on the total photon-photon scattering cross section (ellipticity) and to set an exclusion zone in the mass-inverse coupling parameter plane for scalar/pseudoscalar bosons coupled to two photons.

\section{Apparatus and experimental technique}

\begin{figure}[htb]
\begin{center}
{\includegraphics*[width=18cm]{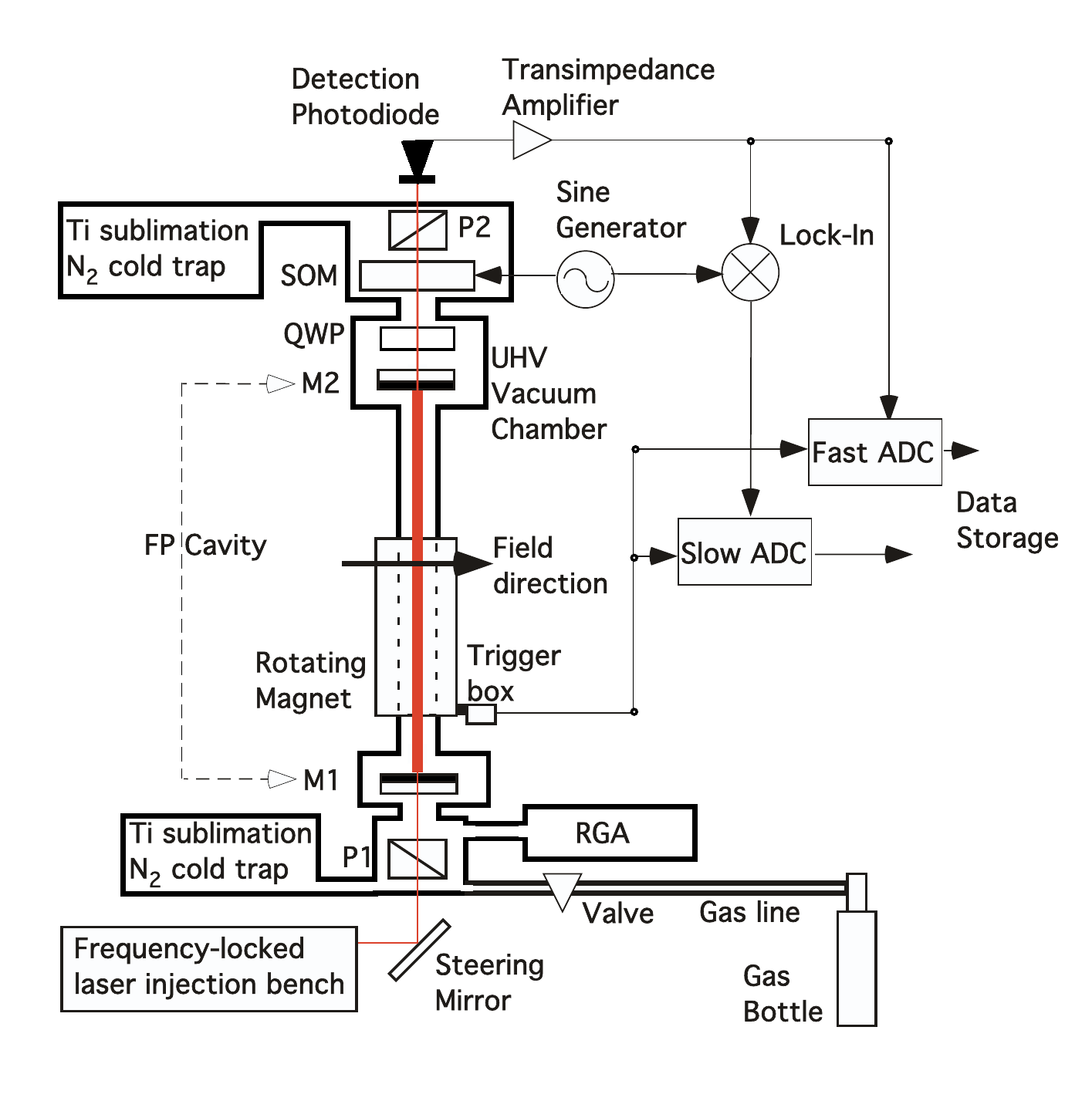}}
\caption{\it Schematic layout of the PVLAS apparatus. See text for description.}
\label{Apparato}
\end{center}
\vspace{-0.6cm}
\end{figure}

Figure \ref{Apparato} shows a schematic layout of the PVLAS apparatus. The set-up develops vertically. Light from a Nd:YAG laser, emitting 800 mW of maximum CW power at 1064 nm, is sent upwards through an ellipsometer consisting of the crossed polarizers P1 and P2.  The ellipsometer encloses a Fabry-Perot optical resonator (FP), formed by a pair of dielectric, multilayer, high reflectivity, 11 m curvature radius mirrors placed 6.4 m apart (M1 and M2),  together with an ellipticity modulator (stress optic modulator, SOM)\cite{Brandi}. A quarter wave plate (QWP) can be inserted between the upper cavity mirror M2 and the SOM in order to make the ellipsometer sensitive to rotations generated within the FP cavity. The resonator (which has negligible diffraction losses) amplifies the optical path in a 1 m long interaction region corresponding to the bore of a dipole superconducting magnet establishing a field lying in the horizontal plane.

A first (lower) granite optical table sits on a concrete platform
mechanically isolated from the rest of the hall floor. The table holds the
laser, various steering mirrors and a Faraday Rotator used to extract the
beam, reflected by the FP, necessary for the feedback system locking the
laser frequency to the cavity frequency \cite{Locking}. It also holds a lower UHV chamber
housing a few optical elements (P1 and M1). A second (upper) granite table
sits on a granite tower fixed onto the same concrete platform as the lower
optical bench. The upper table holds the upper UHV chamber (housing M2,
QWP, SOM and P2). A 4.6 m long, 25~mm diameter, quartz tube, kept under
vacuum and placed vertically, connects the two UHV chambers by traversing the
warm bore of the dipole magnet. The magnet is actually housed within a
cryostat which sits on a 0.9 m radius turntable standing on a concrete beam.
The beam is fastened to the experimental hall floor and spans the concrete
platform hosting the optical tables.  In such a way the magnet assembly is mechanically isolated from the optics. The turntable is actuated by a
low-vibration hydraulic drive and normally rotates the magnet-cryostat
assembly, around a vertical axis, at a frequency $\Omega_{Mag} \approx
0.3$~Hz. The rotation axis is coincident, within a deviation of about 1 mm over 6 m, with the FP cavity axis.

During normal operation the cryostat is filled with liquid He at 4.2 K and the magnet is energized with a current of 2030 A, resulting in a maximum 5.5 T field over the entire interaction region. To allow rotation of the magnet, He feed lines are removed and the coils are shorted and disconnected from the power supply, putting the magnet in persistent current mode. The field intensity, which, due to residual ohmic resistance in the shorting switch, decays at a rate of about 5\% per hour, is then monitored by a set of Hall probes [8]. For analysis purposes, a central value of the field intensity is associated to each data group. 

During ellipsometric measurements in vacuum, the quartz tube and the two main vacuum chambers are kept in vacuum ($P \approx 10^{-8}$ mbar) by two liquid N$_2$ traps combined with Ti sublimation pumps. This pumping scheme has been chosen in order to avoid mechanical
vibrations and possible couplings between the rotating dipole field and ion-pump permanent magnets. The residual gas composition is also monitored by means of a residual gas analyzer (RGA).

To conduct test measurements with high purity gases, a
manifold interface (not shown in Figure \ref{Apparato} for clarity) connects the
lower chamber, through all metal leak valves ("Valve" in Figure \ref{Apparato}) and gas
lines, to several gas bottles. The gas lines can be pumped out up to the
bottle taps.

The light transmitted through the crossed polarizer P2 (analyzer) is detected by a photodiode. The diode current, which contains the physical information, is converted into a voltage by a high-gain (normally $10^{7}$ V/A), low-noise transimpedance amplifier, and then simultaneously acquired by two acquisition chains.
The ``Slow ADC" chain uses a lock-in amplifier, referenced to the same frequency (normally 506 Hz) used to drive the SOM (by the ``Sine Generator" in Figure \ref{Apparato}), to demodulate the input signal so that the interesting peaks will appear as sidebands of the zero frequency. The output of the lock-in is sampled by an ADC gated by trigger signals derived from a series of 32 marks placed on the circumference of the turntable. In this way the angular position of the magnet is always known and absolute signal phases can be determined. The Fast ADC chain directly samples the diode signal at 8.2 kHz (driven by an internal clock) and simultaneously acquires also the trigger signals. Signal phases can then be reconstructed off line. In this chain, which is not demodulated, signals will appear as sidebands of the SOM carrier frequency (506 Hz). ADC outputs of both chains are finally stored for further processing.

\section{Method}
In the PVLAS apparatus signals are detected by measuring the light intensity transmitted by the 
analyzer P2, crossed with P1. An electric field component perpendicular to the entrance polarization fixed by the polarizer P1 may be generated within the FP cavity if an ellipticity $\psi$ is induced by a birefringence or if rotation $\alpha$ is induced by a Faraday effect or a dichroism. A rotation due to the Faraday effect (circular birefringence) is parametrized by the Verdet constant and is linear in the magnetic field intensity $B$. Such a rotation is induced by a magnetic field component parallel to the beam propagation.
If the complex index of refraction is written as $\tilde{n}=n+i\kappa$, where $n$ is the index of 
refraction and $\kappa$ is the extinction coefficient, a dichroism can be described by the 
difference in extinction coefficient $\Delta \kappa=\kappa_{\parallel}-\kappa_{\perp}$ of the medium for two orthogonal 
polarizations, one parallel to some optic axis (in our case the magnetic field) and the other one perpendicular to it. The relationship between the extinction 
coefficient and the absorption coefficient $\mu$ is given by $\mu=4\pi\kappa/\lambda$, where $\lambda$ is the wavelength in vacuum. Similarly, a linear birefringence can be described as the difference between the real indices 
of refraction for the two polarizations $\Delta n=n_{\parallel}-n_{\perp}$. Along a path length $L$, a 
birefringence $\Delta n$ and a dichroism $\Delta \kappa$ generate an ellipticity $\psi$ and a rotation $\alpha$ given by, respectively:
\begin{eqnarray}
\psi=\frac{\pi\Delta n L}{\lambda}\sin2\vartheta
\label{deltaenne}\\
\alpha=\frac{\pi\Delta\kappa L}{\lambda}\sin2\vartheta
\label{deltakappa}
\end{eqnarray}
In the case of a magnetically induced birefringence or dichroism, it is important to note the dependence of both ellipticity and rotation, respectively, on twice the angle $\vartheta$ between the light polarization and the magnetic field.

A phase difference in the electric field of $\pi/2$ between an ellipticity and a rotation allows one to distinguish the two effects. In fact if $\eta(t)$ is the ellipticity induced by the SOM and the QWP is out of the beam path, the intensity $I_{tr}$ transmitted by the analyzer P2 will be
\begin{eqnarray}
I_{tr} = I_{0}\left[\sigma^{2}+\Big|\alpha(t)+\imath\eta(t)+\imath\psi(t)\Big|^{2}\right] \nonumber\\
 =  I_{0}\left[\sigma^{2}+\left(\eta(t)^2+\alpha(t)^{2}+2\psi(t)\eta(t)+\psi(t)^{2}\right)\right]
\label{Itr}
\end{eqnarray}
where $I_{0}$ is the light intensity before the analyzer and $\sigma^2$ is the extinction ratio of the polarizers.
The imaginary nature of the portion of the electric field due to the ellipticities $\psi(t)$ and $\eta(t)$, compared to the real nature of rotations, is explicitly shown. In this experimental condition only the ellipticity will beat with the SOM signal, being then linearized and made detectable.

Therefore, if the magnet rotation is at the angular frequency $\Omega_{Mag}$ and the SOM is modulated at the angular frequency $\omega_{SOM}$, a physical signal generated by a magnetically induced linear birefringence will generate a Fourier component at $ \omega_{SOM}\pm2\Omega_{Mag}$. The factor 2 multiplying $\Omega_{Mag}$ comes from the $2\vartheta$ dependence shown in Eqn.~(\ref{deltaenne}). Residual static ellipticities, always present in a real optical system, are contained in the $\imath\psi(t)$ term of Eqn.~(\ref{Itr}) and can be compensated by acting directly on the SOM itself \cite{Brandi}.

With the QWP inserted the roles of $\psi(t)$ and $\alpha(t)$ will be inverted \cite{Born}: this time the
rotation will appear in Eqn. (\ref{Itr}) as an imaginary number. Furthermore, the QWP may be used in two different orientations by simply exchanging the slow and fast axes. In fact, a real component (rotation) $\alpha(t)$ will become an imaginary one with its sign depending on the QWP orientation. The vector difference of signals measured with the two QWP orientations will isolate rotation effects generated before the QWP. This, in conjunction with the fact that one does not observe signals above background with the FP cavity removed, allows to further narrow down the source of measured rotation or birefringence effects to the cavity alone.

\section{Early observations}
\subsection{Previously published results and relative diagnostic tests}
The results of the rotation measurement from the first series of data taking runs done with the PVLAS apparatus were published in \cite{PVLASrot}. In this letter it was reported the observation of a rotation peak at the frequency $\omega_{SOM}\pm2\Omega_{Mag}$ with an amplitude of $(1.7\pm0.2)\cdot10^{-7}$ rad when 44000 passes are considered, corresponding to $(3.9\pm0.5)\cdot10^{-12}$ rad/pass. The peak appeared with the magnet energized at 5~T and the FP cavity present. Its phase, after an averaging procedure, was found to be compatible with the phase expected from a physical signal. Similar results, albeit with a less clear signature, were found for ellipticity measurements, yielding an average value of $\approx 2\cdot10^{-7}$ for 44000 passes at 5~T \cite{ell_talk}. A first series of diagnostic tests was conducted on these rotation and ellipticity signals with the aim of investigating their nature, physical or instrumental. In the first instance, focus was placed on proving/disproving the fact that the observed peaks were ``optical", meaning that they were present in the spectrum of the detection photodiode current as a consequence of changes of the polarization state of the light propagating through the apparatus. A list summarizing the considered sources of instrumental artifacts, together
with the corresponding experimental tests, is given in Tables \ref{tab:gc-singnal-artefacts1} and \ref{tab:gc-singnal-artefacts2}.

\begin{table}
  \centering
  \caption{List of instrumental artifacts which could account
    for the early observations of rotation and birefringence signals \cite{PVLASrot, ell_talk}. All these sources of artifacts were excluded (see also text).}
  \label{tab:gc-singnal-artefacts1}
  \begin{tabular}{|p{5cm}|p{5cm}|p{6cm}|}\hline
    Origin & Test & Comment \\\hline
     Electronic pick-up (rotation and ellipticity).
    & Measure with field on and the cavity mirrors removed.
    & Pick-up is excluded (see bounds in Table \ref{tab1}).\\\hline
    Mechanical movement due to cryostat rotation.
    & Measure with field off.
    & Effect is excluded (see bounds in Table \ref{tab1}).\\\hline
     Magnetic ro\-tat\-ion\-/ellipticity from a residual gas. 
    & Measure the pressure and composition of the residual gas.
    &  The effect due to the worst
    contaminant is orders of magnitude below the observed effect \cite{nota_tabella}. \\\hline
    Rotation/ellipticity induced by fringe fields on the mirror coatings.
    &Direct measurement of the effect.
    & Magnetically induced rotation and birefringence effects due to fringe fields acting normal and parallel to the mirror surface have been directly measured. Their magnitude cannot account for the observed peaks.\\\hline
        \end{tabular}
\end{table}
\begin{table}
  \centering
  \caption{Table \ref{tab:gc-singnal-artefacts1} continued}\label{tab:gc-singnal-artefacts2} 
  \begin{tabular}{|p{5cm}|p{5cm}|p{6cm}|}\hline
    Origin & Test & Comment \\\hline
    Diffused light from a magnetized inner surface of the cryostat bore (birefringence).
    & Change the geometrical acceptance of the light detection system.
    & A spatial filter is present before the detection
    photodiode. Data taken with several different
    pinhole diameters down to $50\,\mu\mbox{m}$ showed no change in the
    observed signal.\\\hline
    Field-induced movement of the polarizer and/or the QWP (rotation).
    & Measure with field on and the cavity mirrors removed.
    & Excluded by measurements with field on and cavity mirrors
    removed (see comment on pick-ups in Table \ref{tab:gc-singnal-artefacts1}).\\\hline
    Spurious, field-induced ellipticity generated by the SOM modulator.
    & Measure with the field on and the cavity mirrors removed.
    & Excluded by measurements with field on and cavity mirrors
    removed (see comment on pick-ups in Table \ref{tab:gc-singnal-artefacts1}).\\\hline
    Unknown field\--po\-la\-ri\-za\-tion coupling.
    & Eliminate the fringe fields.
    & This coupling cannot come from a \textit{direct} effect of the fringe fields. However, an \textit{indirect} effect, meaning a conspiracy of more than one instrumental artifact, cannot be excluded by the above tests.\\\hline
  \end{tabular}
\end{table}

\subsection{Fringe field effects}
Fringe fields acting on the different optical elements may generate components of both $\psi(t)$ and $\alpha(t)$. These \textit{direct} optical effects were verified not to induce significant instrumental artifacts at twice the rotation frequency of the magnet. The Faraday rotation for the various elements (polarizers, SOM, mirrors and QWP) was measured directly, including the reflective surface of the mirrors. The measured Verdet constants for the mirrors at 1064 nm are $(6.4\pm1.0)\cdot10^{-1}$ rad/T/m for the mirror substrate (fused silica, thickness $8\cdot10^{-3}$ m), and $2\cdot 10^{-7}$ rad/T/reflection for the multilayer high-reflectivity coating. This last number, measured using a test FP cavity, compares well with the results found in \cite{IacopiniFM}. When operating at 5~T the measured vertical stray field component is about $10^{-4}$~T at $\Omega_{Mag}$ and about $10^{-7}$~T at  $2\Omega_{Mag}$. One therefore finds a contribution to the rotation signal amplitude of $1.4\cdot 10^{-6}$ rad at $\Omega_{Mag}$ and of $1.9\cdot 10^{-9}$ at $2\Omega_{Mag}$ (a finesse of 70000 was considered and the presence of both mirrors has been taken into account). It is clear, then, that the typical amplitude of the $\Omega_{Mag}$ rotation signal ($\approx 2-3\cdot 10^{-6}$ rad, see Fig. 2b in \cite{PVLASrot}) is practically entirely due to a fringe field induced Faraday effect, while the contribution at $2\Omega_{Mag}$ is below the observed rotation background. 
In fact, when Helmholtz coils placed around the FP cavity mirrors (see below) are used in feedback mode to cancel all the stray field components including the vertical one, the $\Omega_{Mag}$ signal peak is strongly suppressed. The incomplete suppression can be explained by the fact that the field sensor necessary for the feedback loop is not placed in the exact mirror position, rather, it is fixed at a horizontal distance of about 10 cm.
With respect to the horizontal stray field components, measured to be
$\approx 2.5 \cdot 10^{-4}$ T at $\Omega_{Mag}$ and $\approx 10^{-6}$ T at  $2\Omega_{Mag}$, when using the result reported in \cite{BialolenkerMM}, which give an induced birefringence of $\approx 10^{-13}$ rad/$\mbox{T}^{2}$/reflection, one finds a negligible contribution to the birefringence at $2\Omega_{Mag}$ of $\approx 6\cdot 10^{-12}$. The absence of an effect on the mirrors due to a horizontal field was also verified directly with the Helmholtz coils.
The action of the stray field could however be indirect, meaning that it must couple to some other instrumental effect in order to account for the following empirical findings on the nature of the signal peaks (rotation and ellipticity) reported in \cite{PVLASrot, ell_talk}: the effect is due to the presence of the FP cavity; it changes sign following a change in the orientation of the QWP (rotation) or of the SOM (ellipticity and rotation); there is no measurable direct effect of the stray fields on the cavity mirrors and on the other optical elements.

\subsection{Apparatus upgrades}
With the main intent of reducing the supposed indirect effects of the magnetic fringe fields several upgrades were made to the setup. The laser was changed, going from a Nd:YAG laser at 1064~nm made by Lightwave Inc., to a laser based on the same type of active crystal made by Innolight GmbH. The new laser has actually two beam ports, one emitting at 1064 nm with a maximum power of about 800 mW, and a second one emitting a frequency-doubled beam at 532 nm, with a maximum power of about 100 mW. The 1064 nm beam was used in order to compare data directly with the old measurements. The laser head was also shielded with $\mu$-metal, along with the circuitry used in the electro-optic feedback loop necessary to frequency-lock the laser to the cavity. The previous access structure to the optics tower, which was made almost entirely of iron, was substituted with an aluminum one. All coaxial signal cables were replaced with new cables with better shielding. Two sets of three-axis Helmholtz coils, one set around each cavity mirror, were put in place. They allow both local zeroing of the residual magnetic field and the possibility to actively excite the mirrors with a given field intensity and direction. The initial fixed linear polarization of the light beam has been rotated by $54^{\circ}$ with respect to the previous measurements and is now normal to the beam supporting the rotating magnet.  Finally, a new He gas compressor was installed, increasing the overall efficiency of the magnet cooling cycle and resulting in longer running periods at 4.2 K.

\section{New Results}
Gas measurements, for testing purposes, and vacuum measurements were conducted with the apparatus in the new upgraded configuration. The FP cavity was operating at a typical finesse of 70000. Several diagnostic runs were also done with the Helmholtz coils active or off, in order to test the effect of locally canceling the stray field.
From measurements with field probes when the magnet is energized at 2.3~T, it was found that the stray field is about a factor 50 smaller than at 5~T. Therefore, in order to globally check against fringe field effects,  two measurement campaigns in vacuum were performed with the apparatus in the new upgraded configuration: first both ellipticity and rotation measurements at 2.3~T (no fringe field), then both ellipticity and rotation measurements at a 5~T field (fringe fields present). In addition, a series of diagnostic tests was carried out in order to check whether \textit{indirect} instrumental causes could be used to explain the presence/absence of the rotation and ellipticity signal peaks.

\subsection{Gas test measurements}
\begin{figure}[htb]
\begin{center}
{\includegraphics*[width=15cm]{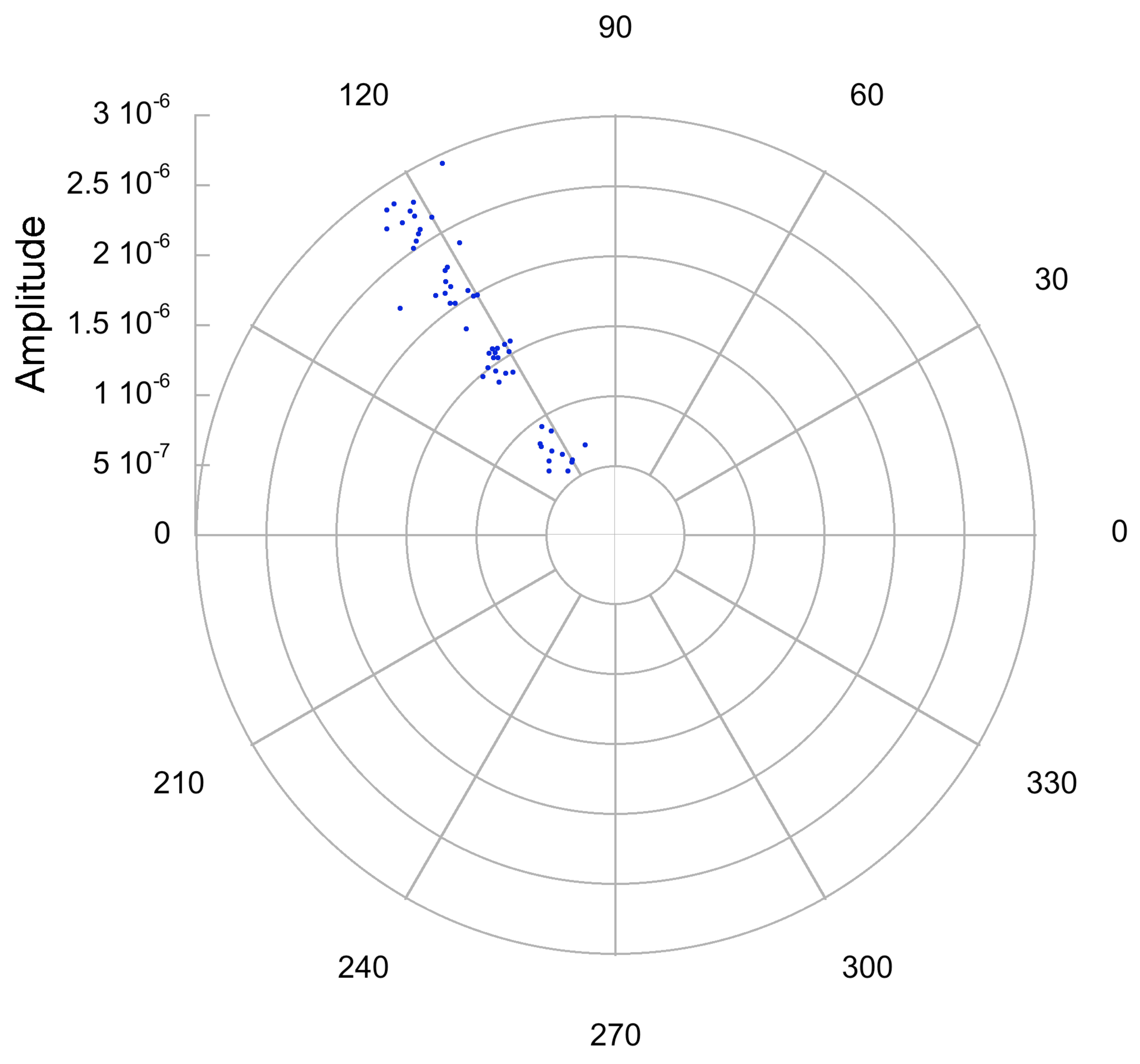}}
\caption{\it Polar plot for the ellipticity signal generated with a 2.3~T magnetic field intensity when Helium gas is present in the vacuum chamber. The figure shows the signal for four different gas pressures: 5, 10, 15 and 20 mbar. Each data point represents amplitude and phase of the signal peak observed in a 100 s long time record. For these data, an ellipticity amplitude of $10^{-6}$ corresponds to a birefringence $\Delta n\approx10^{-17}$}
\label{CM}
\end{center}
\vspace{-0.6cm}
\end{figure}

To verify the correct functioning of the apparatus, test measurements were taken with different gases. In the presence of an external magnetic field, a gas becomes birefringent due to the Cotton-Mouton effect \cite{rizzo}. These measurements also allow checking the physical phase of the Fourier component at twice the rotation frequency of the magnet, $2\Omega_{Mag}$. Indeed, if $\Delta n>0$ the ellipticity is maximum when the angle between the polarization and the slow axis is $45^{\circ}$. In the PVLAS apparatus this translates into a phase at $2\Omega_{Mag}$ of $125^{\circ}$. Figure \ref{CM} shows a polar plot corresponding to the amplitude and phase of the signal due to He gas at four different pressures: 5, 10, 15 and 20 mbar. These measurements were taken with a field intensity of 2.3~T. A gas with a negative birefringence would generate a signal at $180^{\circ}$ with respect to the signals shown in Figure \ref{CM}.
Having defined the physical axis, vacuum results will be presented as components parallel and perpendicular to it. A positive component along the physical axis will mean a positive birefringence.

\subsection{Vacuum measurements}
\begin{figure}[htb]
\begin{center}
{\includegraphics*[width=15cm]{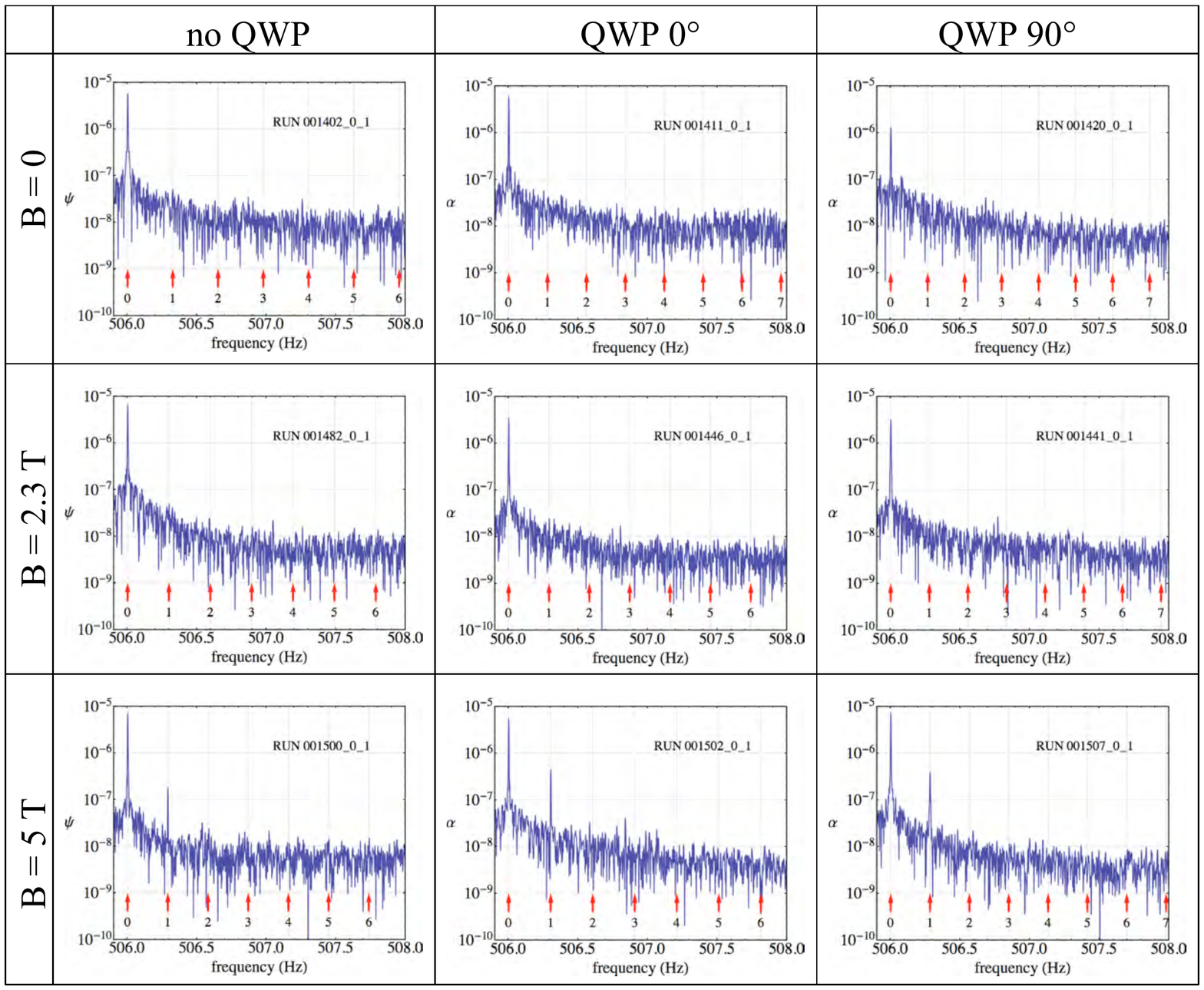}}
\caption{\it Summary table of typical spectra observed in vacuum in the measurements reported here. Each spectrum corresponds to about 600 s of data acquisition time (see text).}
\label{fig2}
\end{center}
\vspace{-0.6cm}
\end{figure}

A summary of the typical spectra obtained in the measurements presented here is shown in Figure \ref{fig2}. A spectrum, corresponding to about 600 s of data acquisition time, is given for each of the three possible configurations of the apparatus and for three different field intensities. The ``no QWP" column shows ellipticity spectra taken with the QWP removed from the beam, and the two columns QWP0 and QWP90 show rotation spectra taken with the QWP in the beam path with two different orientations. The frequency span is chosen in such a way as to show only the upper sidebands of the 506 Hz carrier frequency.

The final results are obtained from the data by taking a vectorial weighted average of 100~s long data subsets. A Fourier transform of the complete data set, for each configuration, is also taken in order to have the best frequency resolution in the possible presence of a peak. Indeed, due to the in-phase data acquisition, a physical signal should occupy a single bin in such a spectrum. 

\subsubsection{Rotation measurements}
In the rotation columns (``QWP0" and ``QWP90") of Figure \ref{fig2}, no signal peaks appear at twice the magnet rotation frequency at 0 T, 2.3 T, and at 5~T. Peaks at twice the magnet rotation frequency remained absent also when the analysis was extended, for a given field intensity, to the entire available data set.

\begin{figure}[htb]
\begin{center}
{\includegraphics*[width=8cm]{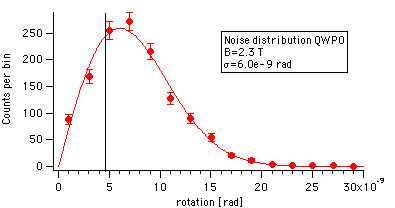}
\includegraphics*[width=8cm]{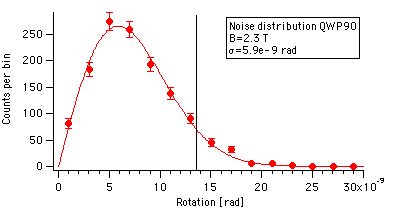}
}
\caption{\it QWP0 and QWP90 noise distributions in the magnet rotation frequency band $1.92\Omega_{Mag}$ -- $2.08\Omega_{Mag}$ for the 2.3~T rotation measurements. The vertical line indicates the resulting amplitude at $2\Omega_{Mag}$ determined from a weighted average of 100~s long data subsets. The value of $\sigma$ for the two configurations is also shown (see text).}
\label{rayleighDiclow}
\end{center}
\vspace{-0.6cm}
\end{figure}

Histograms of the noise from the Fourier spectrum for the QWP0 and QWP90 data, in the frequency interval $1.92\Omega_{Mag} - 2.08\Omega_{Mag}$, are shown in Figure \ref{rayleighDiclow} for the 2.3~T field intensity. A fit with a Rayleigh probability distribution is superimposed. This is the probability density function which results for the amplitude of a signal having a Gaussian distribution along two orthogonal axes with equal standard deviations $\sigma$. A vertical line indicates the values obtained from the weighted average of the 100~s subdatasets at $2\Omega_{Mag}$. The components of these vectors projected along the physical axis and in the direction normal to it are given in Table \ref{tab_2.3T_comps}.

\begin{table}
\begin{tabular}{|p{4cm}|p{4cm}|p{4cm}|}
\hline\hline
Rotation vector at 2.3~T & Component parallel to physical axis & Component normal to physical axis\\\hline
QWP0 & $(-3.5\pm6.0)\cdot10^{-9}$ rad & $(-2.9\pm6.0)\cdot10^{-9}$ rad\\
QWP90 & $(4.6\pm5.9)\cdot10^{-9}$ rad & $(-13\pm5.9)\cdot10^{-9}$rad \\
\hline\hline
\end{tabular}
\caption{Components of the QWP0 and QWP90 vectors at 2.3~T (see text).}
\label{tab_2.3T_comps}
\end{table}

It is evident from Figure \ref{rayleighDiclow} that neither the QWP0 nor the QWP90 data present a peak above the noise.
From the Rayleigh cumulative probability distribution ($F(x)=1-e^{-0.5(\frac{x}{\sigma})^2}$) one can give a limit on the induced rotation at $2\Omega_{Mag}$ of $\alpha_{QWP0}\leq1.5\cdot10^{-8}$~rad at a 95\% confidence level in the QWP0 configuration, and $\alpha_{QWP90}\leq1.4\cdot10^{-8}$~rad at a 95~\% confidence level in the QWP90 configuration. By taking the vector average $\Delta=\frac{QWP0-QWP90}{2}$ between the QWP0 and QWP90 results, where the minus sign takes into account
the fact that the two measurements should have different signs, one obtains an amplitude $\alpha_{2.3T}=(6.5\pm4.2)\cdot10^{-9}$~rad. Interpreting the value of the uncertainty as the standard deviation of a Rayleigh distribution one can give a limit on rotation at 2.3~T of $1.0\cdot10^{-8}$~rad at a 95\% confidence level. The total measurement time at 2.3~T field intensity was 47300 s.

\begin{figure}[htb]
\begin{center}
{\includegraphics*[width=8cm]{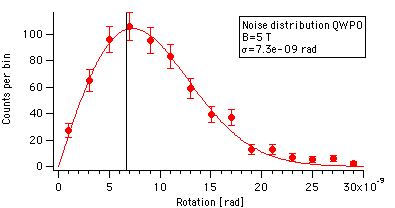}
\includegraphics*[width=8cm]{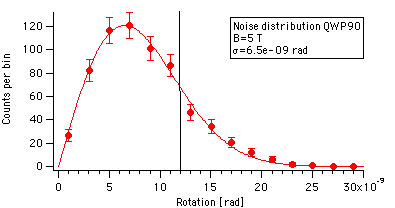}
}
\caption{\it QWP0 and QWP90 noise distributions in the magnet rotation frequency band $1.92\Omega_{Mag}$ -- $2.08\Omega_{Mag}$ for the 5~T rotation measurements. The vertical line indicates the resulting amplitude at $2\Omega_{Mag}$ determined from a weighted average of 100~s long data subsets. The value of $\sigma$ for the two configurations is also shown (see text).}
\label{rayleighDichigh}
\end{center}
\vspace{-0.6cm}
\end{figure}

The corresponding noise histograms for the 5~T measurements are shown in Figure \ref{rayleighDichigh} in the frequency band $1.92\Omega_{Mag}$ --- $2.08\Omega_{Mag}$. The weighted vector averages of the 100~s data subsets results are presented in Table \ref{tab_5.5T_comps}, where, as before, the components of the QWP0 and QWP90 vectors projected along the physical axis and in the direction normal to it are given.

\begin{table}
\begin{tabular}{|p{4cm}|p{4cm}|p{4cm}|}
\hline\hline
Rotation vector at 5~T & Component parallel to physical axis & Component normal to physical axis\\\hline
QWP0 & $(2.5\pm7.3)\cdot10^{-9}$ rad & $ (6.2\pm7.3)\cdot10^{-9}$ rad\\
QWP90 & $(2.1\pm6.5)\cdot10^{-9}$ rad & $(-12\pm6.5)\cdot10^{-9}$ rad\\
\hline\hline
\end{tabular}
\caption{Components of the QWP0 and QWP90 vectors at 5~T (see text).}
\label{tab_5.5T_comps}
\end{table}

By taking a vectorial average one obtains $\alpha_{5T}=(9.1\pm4.9)\cdot10^{-9}$, again well within the 95\% confidence limit of $1.2\cdot10^{-8}$ rad. The total integration time at 5~T was 30100 seconds.

As can be seen in Figure \ref{fig2}, a peak appears at the magnet rotation frequency $\Omega_{Mag}$ when working at 5~T, and this is interpreted as due to a Faraday rotation in the FP cavity mirrors caused by the fringe field vertical component (see discussion above).

\subsubsection{Ellipticity measurements}
In the ellipticity column of Figure \ref{fig2} no signal peaks appear at $2\Omega_{Mag}$ at 0~T, 2.3~T, and at 5~T. 
However, a small peak at $2\Omega_{Mag}$ appeared in the 5~T data when the analysis was extended to the entire data set. The peak at the magnet rotation frequency $\Omega_{Mag}$ present in the 5~T row of Fig. \ref{fig2} can be interpreted partly as due to the mirror Faraday rotation transformed into an ellipticity by the presence of the FP cavity itself \cite{GuidoCav} and partly to beam movements on the cavity mirrors. In fact, dielectric mirrors present an ordered birefringence ``map" which has a gradient \cite{micossi}. A beam movement at a given frequency $\Omega$ will therefore generate an ellipticity at the same frequency. This effect has been measured yielding an ellipticity gradient of $\approx 10^{-6} \mu \mbox{m}^{-1}$.

\begin{figure}[htb]
\begin{center}
{\includegraphics*[width=8cm]{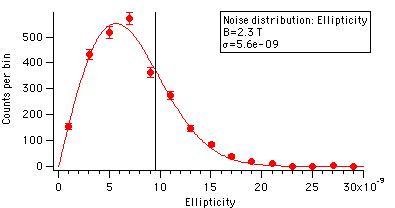}
\includegraphics*[width=8cm]{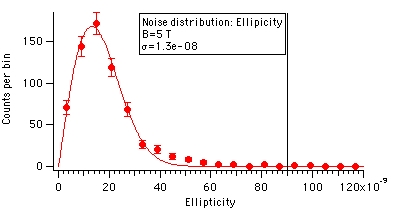}
}
\caption{\it Noise distributions in the magnet rotation frequency band $1.92\Omega_{Mag}$ -- $2.08\Omega_{Mag}$ for the 2.3~T (left) and 5~T (right) ellipticity measurements. The vertical line indicates the resulting amplitude $2\Omega_{Mag}$ determined from a weighted average of 100~s long data subsets. The value of $\sigma$ for the two field intensities is also shown (see text).}
\label{rayleighBirlow}
\end{center}
\vspace{-0.6cm}
\end{figure}

A histogram of the noise between $1.92\Omega_{Mag}$ and $2.08\Omega_{Mag}$ is shown in Figure \ref{rayleighBirlow} at left for the 2.3~T data, with the vertical line indicating the value determined at $2\Omega_{Mag}$ as the weighted average of 100~s data subsets. The resulting amplitude is $9.5\cdot10^{-9}$. By using the $\sigma$ obtained from the Rayleigh distribution as an estimate of the error on the measured amplitude, the value at $2\Omega_{Mag}$ is well within the 95\% confidence limit. An upper limit of $\psi_{2.3T} \leq 1.4\cdot10^{-8}$ at a 95~\% confidence level can therefore be determined from the ellipticity data at 2.3~T.


At 5~T the ellipticity measurements show a peak at $2\Omega_{Mag}$ (Fig. \ref{rayleighBirlow} at right). The amplitude of the peak at a central field of 5~T is $\psi_{5T}=(9.0\pm 0.9)\cdot 10^{-8}$, well above background. By considering a $B^2$ dependence of this possible physical signal, at 2.3~T one would have expected an ellipticity $\psi_{exp}=1.9\cdot10^{-8}$ which, given a $\sigma$ of $5.6\cdot10^{-9}$ at 2.3~T, is excluded at better than a $99\%$ confidence limit.  We conclude that the ellipticity peak at 5~T must be therefore of instrumental origin. Integration time at 5~T was 14300 s.

\subsubsection{Summary of results}
Table \ref{tab1} gives the $95\%$ confidence level background values for both rotation and ellipticity measurements. Data were taken with a typical cavity finesse of 70000, corresponding to about 45000 passes through the magnetic field zone.
The total measurement time at the 2.3~T field intensity was 47300 s for rotation and 65200 s for ellipticity, while at 5~T it was 30100~s. 
\begin{table}
\begin{tabular}{ccc}
\hline\hline
Meas. Type & 2.3~T & 5~T\\\hline
Rotation & $1.0\cdot 10^{-8}$ rad & $1.2\cdot 10^{-8}$ rad\\
Ellipticity & $1.4\cdot 10^{-8}$ & \\
\hline\hline
\end{tabular}
\caption{Measured rotation and ellipticity backgrounds (95\% c.l.) at two magnetic field intensities.}
\label{tab1}
\end{table}

No signal peaks were observed in the trasmitted intensity spectra at twice the magnet rotation frequency both in rotation and ellipticity at 2.3 T. Assuming a $B^2$ dependence of the previously published rotation signal ($1.7\cdot 10^{-7}$ rad at 5~T with 44000 passes in the cavity \cite{PVLASrot}), one should expect to observe, at 2.3~T, a rotation peak with an amplitude of $3.6\cdot 10^{-8}$ rad. Since the $\sigma$ of the 2.3~T rotation measurement is about one order of magnitude smaller than this value, such a signal can be excluded with a very high confidence level. This fact immediately excludes a possible $B^2$ dependence of the published rotation signal. Furthermore, the absence of rotation peaks in the 5~T data directly contradicts the observations published in \cite{PVLASrot}. In this latter work, the relatively large dispersion of the data was treated under the hypothesis of an underlying Gaussian cause for the variability, resulting in an error estimate which, in view of the present results, was probably too small.

\subsection{Diagnostic tests on \textit{indirect} instrumental artifacts}
The vacuum measurement runs with the magnet cold and energized were followed by a series of tests where it was attempted to induce ellipticity/rotation signals by acting externally on possible sources of indirect coupling to the light polarization. Tables \ref{table4} and \ref{table4cont} give a summary of these tests along with the relevant comments. As a general comment, one observes here that three of the sources which were investigated could potentially cause signals in both birefringence and rotation at the frequency $2\Omega_{Mag}$. However, when an attempt was made to stimulate these sources with local magnetic fields of a few gauss (as the fringe fields generated by the superconducting magnet) the measured effects were smaller by a factor of about 10 than the vacuum effects reported in \cite{PVLASrot}. To check against the possibility that these vacuum signals arise as a combination of artifact sources such as those listed in Tables \ref{tab:gc-singnal-artefacts1} and \ref{table4}, further tests were conducted by attempting to simultaneously excite two of these sources. Table \ref{table5} presents a short summary of these last tests. Also in this case none of the combinations of effects which were investigated could account for the results reported in \cite{PVLASrot, ell_talk}.

\begin{table}
  \centering
  \caption{List of possible sources of \textit{indirect} coupling to the light polarization}\label{tab:indirect_artifacts} 
  \begin{tabular}{|p{5cm}|p{5cm}|p{6cm}|}\hline
    Source & Test & Comment \\\hline
    Fringe-field induced modulation of the frequency-locking circuit offset (ellipticity and rotation).
    & Directly modulate with a signal the locking circuit offset voltage.
    &  Can generate both a rotation and an ellipticity at the same frequency of the modulation.\\\hline
    Fringe-field induced amplitude modulation of the SOM carrier signal.
    & Modulate the amplitude of the sine-wave signal exciting the SOM (typical residual modulation in actual running conditions is $\leq 10^{-3}$).
    & Can generate a signal at the same frequency of the modulation. Can generate a signal at the second harmonic of the modulation frequency if modulated deeply enough. Cannot be excited by a local field of the order of a few gauss.\\\hline
    Fringe-field induced amplitude modulation of the laser intensity (ellipticity and rotation).
    & Modulate the supply current of one of the laser pump diodes (typical residual modulation in actual running conditions is $\leq 10^{-3}$).
    & Can generate a signal at the same frequency of the modulation. Can generate a signal at the second harmonic of the modulation frequency if modulated deeply enough. Cannot be excited by a local field of the order of a few gauss.\\\hline 
   \end{tabular}
  \label{table4}
\end{table} 
    
\begin{table}
 \centering
\caption{Table \ref{tab:indirect_artifacts} continued}\label{tab:indirect_artifacts2} 
\begin{tabular}{|p{5cm}|p{5cm}|p{6cm}|}\hline
    Fringe-field action on injection bench Faraday rotator (ellipticity and rotation).
    & Use an external Helmholtz coil to create a controlled local field of a few gauss modulated at a given frequency $\Omega$.
    & An amplitude modulation at $\Omega$ and at $2\Omega$ is observed in the light intensity transmitted through the analyzer P2 (see Figure \ref{Apparato}). Signals at $\Omega$ and at $2\Omega$ are present. The ratios of the amplitudes at $\Omega$ and at $2\Omega$ are the same for both the amplitude and the ellipticity/rotation modulations. This indicates a $\Omega$-$2\Omega$ correlation which is absent in the vacuum data of \cite{PVLASrot}. The amplitudes of the signals thus generated are also a factor 10 smaller than those reported in \cite{PVLASrot}. \\\hline
   Residual mechanical movements (ellipticity).
    & Modulate by periodically moving a 40 kg inertial mass placed on the upper optical bench.
    & Can generate an ellipticity at the modulation frequency.\\\hline
  \end{tabular}
  \label{table4cont}
\end{table}

\begin{table}
  \centering
  \caption{Tests combining possible \textit{direct} ad \textit{indirect} sources of instrumental artifacts}\label{tab:crossed_artifacts} 
  \begin{tabular}{|p{5cm}|p{5cm}|p{6cm}|}\hline
    Source & Test & Comment \\\hline
    Fringe-field induced modulation of the frequency-locking circuit offset (ellipticity and rotation) combined with a Faraday fringe field on the cavity upper mirror (rotation).
    & Use external Helmholtz coils to create local fields. Modulate both fields at the same frequency.
    &  A rotation and/or a ellipticity can be generated at the modulation frequency. However, there appears no effect at the sum frequency (twice the modulation frequency).\\\hline
    Fringe-field induced amplitude modulation of the SOM (ellipticity) combined with a Faraday fringe field on the cavity upper mirror (rotation).
    & Use an external Helmholtz coil to create the local Faraday field and modulate at some frequency. Modulate the amplitude of the sine-wave signal exciting the SOM at the same frequency.
    & A rotation and/or a ellipticity can be generated at the modulation frequency. There appears no effect at the sum frequency (twice the modulation frequency) if the residual modulation on the SOM is the same as in actual running conditions ($\leq 10^{-3}$).\\\hline
    Fringe-field induced excitation of the Faraday Rotator  (ellipticity and rotation) combined with a Faraday fringe field on the cavity upper mirror (rotation).
   & Use external Helmholtz coils to create local fields. Modulate both fields ay the same frequency.
  & A rotation and/or a ellipticity can be generated at the modulation frequency. The amplitude of this effect is roughly the in-phase sum of the two excitations. There appears to be no additional effect at the sum frequency.\\\hline 
  Fringe-field induced excitation of the Faraday Rotator  (ellipticity and rotation) combined with residual mechanical movements (ellipticity).
   & Use external Helmholtz coils to create local field on Faraday rotator. Periodically move a 40 kg inertial mass placed on the upper optical bench. Modulate both excitations at the same frequency.
    & A rotation and/or a ellipticity can be generated at the modulation frequency. The amplitude of this effect is roughly the in-phase sum of the two excitations. There appears to be no additional effect at the sum frequency.\\\hline
  \end{tabular}
  \label{table5}
\end{table}

\section{Discussion and conclusions}
The rotation measurements done at a field intensity of 5~T indicate that the rotation signal reported in \cite{PVLASrot} was due to an instrumental artifact. Furthermore, the 2.3 T measurements, where no signal peak is visible both in rotation and in ellipticity, render improbable the hypothesis that the apparatus upgrades have themselves introduced an instrumental artifact exactly canceling the ``true" previous signal, including the $B^{2}$ dependence. Recent direct measurements done using the photon regeneration scheme confirm these conclusions \cite{rizzoGammeV}.
The limiting observed background values for rotation and ellipticity are, respectively:
\begin{eqnarray}
\alpha \leq 2.7\cdot10^{-13}~\mbox{rad/pass at 95\% c.l. at 5~T}
\label{limits1}\\
\psi \leq 3.1\cdot10^{-13} ~\mbox{1/pass at 95\% c.l. at 2.3 T}
\label{limits2}
\end{eqnarray}
The rotation limit is calculated by combining the QWP0 and QWP90 data, that is by taking the semi-difference of the weighted averages of the two data sets.
These figures, using Eqns. \ref{deltaenne} and \ref{deltakappa}, also set limits on the values of the observed magnetically induced birefringence and dichroism of vacuum:

\begin{eqnarray}
\Delta n \leq 1.1\cdot10^{-19}~\mbox{at 2.3 T}\\
\Delta \kappa \leq 0.9\cdot10^{-19}~\mbox{at 5~T}
\end{eqnarray}

This last value corresponds to a difference in the absorption coefficients for the two orthogonal polarizations of $\Delta \mu \leq 1.1 \cdot 10^{-14}~\mbox{cm}^{-1}.$
\begin{figure}[htb]
\begin{center}
{\includegraphics*[width=18cm]{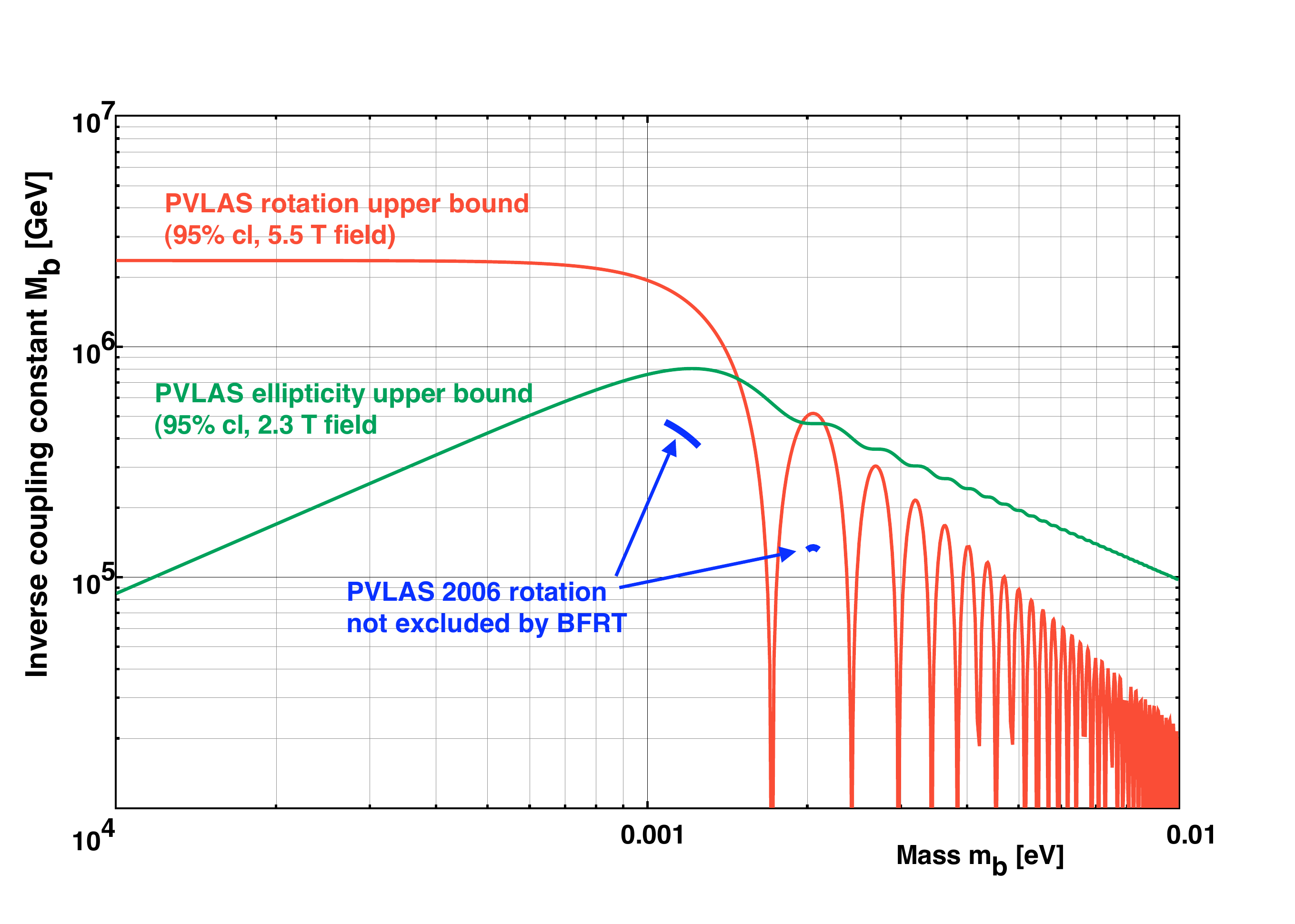}}
\caption{\it Upper bounds on mass and inverse coupling constant for scalar/pseudoscalar bosons coupled to two photons. These bounds are derived from the background values reported in Table \ref{tab1} taking into account  45000 passes in the FP cavity. Also shown are the regions calculated from the data published in \cite{PVLASrot} and compatible with the bounds reported in \cite{Cameron}. The new data completely exclude the previous 2006 results \cite{PVLASrot}.}
\label{fig3}
\end{center}
\vspace{-0.6cm}
\end{figure}

Furthermore, the limiting values for observed rotation  and ellipticity can be used to draw exclusion zones in the mass-inverse coupling plane for light, neutral bosons coupling to two photons \cite{Maiani, Raffelt, Mass˜}. Figure \ref{fig3} shows a plot of such a parameter space. The plot contains curves calculated from the figures given in Eqns. \ref{limits1} and \ref{limits2}, taking into account 45000 passes in the interaction region, and shows the two portions of parameters space resulting from the previously observed rotation signal \cite{PVLASrot} and not excluded by the BFRT results \cite{Cameron}.
Finally, the ellipticity figure can be used to set an upper bound on the total photon-photon cross section \cite{Duane}, of $\sigma_{\gamma\gamma} < 4.5 \cdot 10^{-34}$ barn.

\section{Acknowledgements}
We wish to thank S. Marigo, A. Zanetti and G. Venier for their invaluable and precious technical help on the construction and running of the apparatus. We also acknowledge all the people of the Laboratori Nazionali di Legnaro that helped us and made this experiment possible.


\begin{thebibliography}{00}
\bibitem{nota_tabella}
  The transverse component of the magnetic field gives a Cotton-Mouton effect (birefringence) \cite{rizzo}, while the longitudinal component is responsible for a Faraday effect (rotation) \cite{IacopiniFM}.
\end{thebibliography}
\end{document}